\documentclass[%
 reprint,
 amsmath,amssymb,
 aps,
 floatfix
]{revtex4-2}

\usepackage{graphicx}
\usepackage{dcolumn}
\usepackage{bm}
\usepackage{braket}
\usepackage{array}
\usepackage{amsmath}
\usepackage{color}
\usepackage{float}
\usepackage{placeins}
\usepackage{here}
\usepackage{algorithmic}
\usepackage{graphicx}
\usepackage{textcomp}

\usepackage{braket}

\DeclareMathOperator*{\argmax}{arg\,max}

\begin{document}

\title{Investigation of Automated Design of Quantum Circuits \\ for Imaginary Time Evolution Methods \\ Using Deep Reinforcement Learning}

\author{Ryo Suzuki}
\email{al20108@shibaura-it.ac.jp}
\author{Shohei Watabe}
\email{watabe@shibaura-it.ac.jp}

\affiliation{Graduate School of Engineering and Science, Shibaura Institute of Technology, Toyosu 3-7-5, Tokyo 135-8548, Japan}

\begin{abstract}
Efficient ground state search is fundamental to advancing combinatorial optimization problems and quantum chemistry. While the Variational Imaginary Time Evolution (VITE) method offers a useful alternative to Variational Quantum Eigensolver (VQE), and Quantum Approximate Optimization Algorithm (QAOA), its implementation on Noisy Intermediate-Scale Quantum (NISQ) devices is severely limited by the gate counts and depth of manually designed ansatz. Here, we present an automated framework for VITE circuit design using Double Deep-Q Networks (DDQN). Our approach treats circuit construction as a multi-objective optimization problem, simultaneously minimizing energy expectation values and optimizing circuit complexity. By introducing adoptive thresholds, we demonstrate significant hardware overhead reductions. In Max-Cut problems, our agent autonomously discovered circuits with approximately 37\% fewer gates and 43\% less depth than standard hardware-efficient ansatz on average. For molecular hydrogen ($H_2$), the DDQN also achieved the Full-CI limit, with maintaining a significantly shallower circuit. These results suggest that deep reinforcement learning can be helpful to find non-intuitive, optimal circuit structures, providing a pathway toward efficient, hardware-aware quantum algorithm design. 
\end{abstract}

\maketitle

\section{Introduction}

Noisy Intermediate-Scale Quantum (NISQ) computers~\cite{preskill2018quantum} are fundamentally constrained by decoherence and limited gate fidelities. Consequently, the practical utility of NISQ algorithms depends on the ability to design shallow quantum circuits that minimize gate counts while maintaining computational accuracy. For applications such as combinatorial optimization problems and quantum chemical calculations, determining the optimal gate sequence and circuit topology is a non-trivial task that typically demands extensive domain expertise and manual heuristic tuning. As the performance of NISQ algorithms is highly sensitive to the underlying ansatz structure, there is a critical need for automated, hardware-aware circuit design methodologies. Recently, reinforcement learning (RL) has emerged as a promising candidate for self-learning autonomous circuit design~\cite{mateusz, fosel, kolle2024reinforcement, kimura2022quantum}.

Earlier studies have explored various RL architectures for circuit optimization. Mateusz et al.~\cite{mateusz} employed a Double Deep-Q Network (DDQN) to design ansatz for the Variational Quantum Eigensolver (VQE)~\cite{cerezo2021variational}. 
While their framework successfully derived the ground state of lithium hydride (${\rm LiH}$) with fewer gates than standard Hardware-Efficient ansatz~\cite{kandala2017hardware}, the reward structure lacks an explicit incentive for reducing circuit size, providing only a negative penalty for exceeding depth constraints. 
Similarly, Fösel et al.~\cite{fosel} utilized the RL based on Proximal Policy Optimization (PPO)~\cite{schulman2017proximal} to optimize pre-existing inefficient circuits. Although they achieved a 27\% reduction in depth for 12-qubit systems, the optimization efficiency remained highly dependent on the quality of the initial circuit, such as randomly generated circuits or initial circuits from the Quantum Approximate Optimization Algorithm (QAOA)\cite{farhi2014quantum}. 
Furthermore, Kolle et al.~\cite{kolle2024reinforcement} applied Markov Decision Processes (MDPs) for state preparation but observed a tendency toward structural redundancy as the target-state complexity increased.

Despite these advancements, most RL-based automation for the NISQ algorithms has focused on VQE and QAOA. The Variational Imaginary Time Evolution (VITE) method~\cite{motta2020determining}---a powerful technique for ground-state exploration and generating probability distributions for machine learning---remains relatively under-explored in the context of automated ansatz design. Moreover, existing RL frameworks for VQE and QAOA lack an explicit reward incentive for circuit-size reduction. This absence not only makes it difficult to navigate the trade-off between energy and compactness but also fails to provide any guarantee that the resulting circuit is structurally optimal, often leading to redundant, excessively deep architectures that are impractical for noisy hardware.

We here study an automated design framework of quantum circuits for the VITE method using a DDQN architecture that simultaneously  optimizes the circuit to reach the ground state while ensuring its structural complexity is minimized. 
We evaluate the performance of our method on the Max-Cut problem and molecular ground-state simulations. 
In both cases, we observed a tendency for the expectation value of the Hamiltonian and circuit depths to decrease. 
However, in quantum chemistry calculations, the performance is sensitive to the reward and threshold, and the energy expectation value mainly converges to the Hartree-Fock approximation value rather than the exact solution (Full-CI) for a suboptimal threshold. 
To resolve this issue, we implemented an adoptive thresholding mechanism for the expectation value of the Hamiltonian. 
By using this mechanism, we significantly improved the success rate of discovering shallow circuits capable of reaching the Full-CI value. Furthermore, by analyzing various circuits generated by the DDQN, we extracted skelton structures from the designed circuits, demonstrating the potential for further reduction in the number of gates. 
These findings not only demonstrate the efficacy of RL in non-unitary algorithm design but also provide a potential foundation for establishing systematic guidelines for future, hardware-aware ansatz construction.


\section{Variational Imaginary Time Evolution (VITE) Method}

The imaginary time evolution (ITE) method~\cite{motta2020determining, yuan2019theory} is a powerful NISQ-compatible algorithm designed for ground-state preparation. Unlike real-time evolution, where the time evolution of quantum systems is described by the Schrödinger equation:
\begin{equation}
 i\frac{\partial \ket{\psi(t)}}{\partial t} = H\ket{\psi(t)}, 
\end{equation}
the ITE evolves the system according to the Schr\"{o}dinger equation transformed via a Wick rotation~\cite{wick1954properties}, where the real time $t$ is replaced by the imaginary time $\tau = it$, given by  
\begin{equation}
 \frac{\partial \ket{\psi(\tau)}}{\partial \tau} = -(H-E_{\tau})\ket{\psi(\tau)}. 
 \label{SchroEqITE}
\end{equation}
Here, $ E_{\tau} = \braket{\psi(\tau)|H|\psi(\tau)}$ represents the energy expectation value at the imaginary-time $\tau$ to conserve the norm of the quantum state vector. 
The quantum state $\ket{\psi(\tau)}$ satisfying Eq.~\eqref{SchroEqITE} is given by 
\begin{equation}
 \ket{\psi(\tau)} = \frac{e^{-H\tau}\ket{\psi(0)}}{\sqrt{\braket{\psi(0)|e^{-2H\tau}|\psi(0)}}}, 
\end{equation}
where an initial state with nonzero overlap with the ground state converges to the ground state of the Hamiltonian through a long imaginary time evolution ($\tau\to\infty$). 

Since the ITE is a non-unitary process, it cannot be directly implemented on a quantum gate-based computer. To overcome this, the McLachlan's variational principle \cite{mclachlan1964variational, broeckhove1988equivalence} is employed in VITE to approximate the evolution using a parameterized trial state $\ket{\phi(\boldsymbol{\theta}(\tau))}$, where $\boldsymbol{\theta}(\tau) = (\theta_1(\tau), \theta_2(\tau), \dots, \theta_N(\tau))^T$ represents a set of time-dependent rotation angles in the ansatz. The McLachlan's principle minimizes the distance between the exact ITE and the constrained evolution within the parameterized manifold:
\begin{align}
 \delta \left | \left( \frac{\partial}{\partial \tau} + H - E_\tau \right) \ket{\phi(\tau)} \right | = 0. 
\end{align}
This optimization provides linear equations governing the time evolution of the parameters $\boldsymbol{\theta}$:
\begin{equation} \label{eq:parameter_evolution}
 \sum_j A_{ij} \dot{\theta}_j = C_i,
\end{equation}
where $\dot{\theta}_j$ denotes the time derivative of $\theta_j$. The coefficients $A_{ij}$ and $C_i$ are respectively given by 
\begin{align}
 A_{ij} &= \text{Re} \left( \frac{\partial \bra{\phi}}{\partial \theta_i} \frac{\partial \ket{\phi}}{\partial \theta_j} \right), \\
 C_i &= \text{Re} \left( -\sum_\alpha \lambda_\alpha \frac{\partial \bra{\phi}}{\partial \theta_i} P_\alpha \ket{\phi} \right).
\end{align}
Here, $P_\alpha$ and $\lambda_\alpha$ represent the Pauli operators and their corresponding coefficients in the system Hamiltonian $H = \sum_\alpha \lambda_\alpha P_\alpha$. By numerically solving Eq.~\eqref{eq:parameter_evolution} on a classical computer, the parameters $\boldsymbol{\theta}$ are updated iteratively, driving the trial state toward the ground state. This framework enables the approximation of non-unitary ITE through a sequence of unitary operations via continuous parameter updates.

\section{Reinforcement Learning}
\begin{figure}[t]
	\centering
    \includegraphics[keepaspectratio, width=87mm]{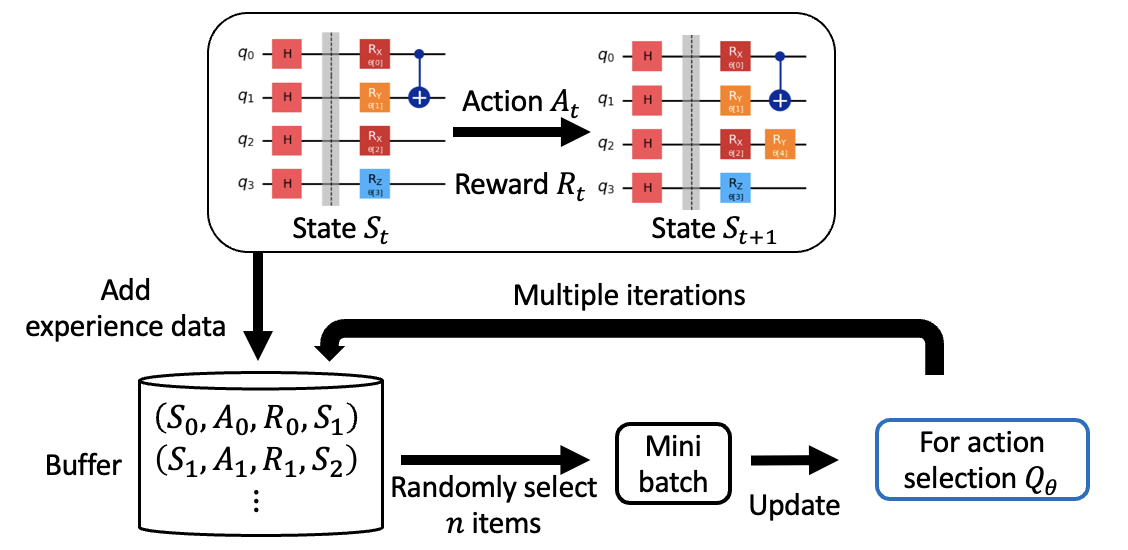}
	\caption{Schematics of the framework and the process of experience replay when applied to quantum circuit design.}
	\label{exp_replay2}
\end{figure}

Reinforcement learning (RL)~\cite{kaelbling1996reinforcement} is a computational approach where an autonomous agent learns to make sequences of decisions to maximize a cumulative reward within a given environment. The learning process relies on trial-and-error interactions, enabling the agent to optimize its strategy (policy) without pre-existing labeled datasets. In our framework, each interaction is represented by a transition tuple consisting of four fundamental elements (see also Fig.~\ref{exp_replay2}):
\begin{itemize}
  \item \textbf{State} $S_t$: The current configuration of the environment. In this study, $S_t$ represents the current structure of the quantum circuit.
  \item \textbf{Action} $A_t$: The operation performed by the agent. We define the action space as the addition of one quantum gate from a gate set $\{{R_x, R_y, R_z, I}, \text{CNOT}\}$.
  \item \textbf{Reward} $R_t$: A scalar feedback signal evaluating the performance. Here, $R_t$ is determined by both the energy expectation value and the circuit complexity.
  \item \textbf{Next State} $S_{t+1}$: The state resulting from taking action $A_t$ in state $S_t$, corresponding to the updated circuit structure.
\end{itemize}

A single interaction cycle, moving from $S_t$ to $S_{t+1}$ and receiving $R_t$, is defined as one \textit{step}. An \textit{episode} encompasses the entire sequence from the initial circuit to the final design. 
The agent's ultimate objective is to discover an optimal sequence of actions that yields a high-precision, low-depth quantum circuit.

\subsection{$\varepsilon$-greedy}

Balancing exploration and exploitation is crucial in reinforcement learning. 
Exploration enables the agent to traverse diverse circuit configurations, whereas exploitation leverages its experience to converge toward an optimal policy.
To manage this trade-off, we employ an $\varepsilon$-greedy method~\cite{tokic2010adaptive}, where the agent selects a random action with probability $\varepsilon$ and the action that maximizes the Q-value with probability $1-\varepsilon$.

Typically, $\varepsilon$ is initialized to a high value to prioritize exploration and is decayed toward a lower bound as learning progresses. A standard approach for decaying the exploration probability is given by $\varepsilon = \Gamma^t$, 
where $\Gamma$ is a decay rate ($0 < \Gamma < 1$), and $t$ represents the cumulative number of steps across the entire training process. 
However, in the current task, such a global decay tends to reduce the exploration rate prematurely. This often results in highly constrained exploration of novel circuit configurations during the later stages of the learning process.

To maintain a consistent exploration-exploitation trade-off throughout the training progress regardless of the cumulative number of steps, 
we employ a modified decay function: 
\begin{equation}
\varepsilon = \Gamma^{(\nu-1) + (t_\nu-1)},
\label{eq:e_origin}
\end{equation}
which can show a gradual transition from broad exploration to stable exploitation as the episode progresses. 
Here, $\nu$ denotes the episode index and $t_\nu$ represents the step count within the $\nu$-th episode. 
By incorporating the episode index $\nu$ and the step count $t_\nu$ within each episode—rather than the cumulative steps across the entire training process—the decay function ensures that the exploration rate scales with the overall learning progress while preserving sufficient capacity for trial-and-error.

\subsection{Double Deep-Q Network (DDQN)}

To facilitate efficient and stable circuit design, we employ the Double Deep-Q Network (DDQN) algorithm \cite{van2016deep}, a variant of deep reinforcement learning that enhances the quality of action selection (exploitation). In this framework, the agent estimates the action-value function, or Q-value $Q(S_t, A_t)$, which represents the expected cumulative reward obtained by taking action $A_t$ in state $S_t$ and following the optimal policy thereafter.
A critical challenge in standard Deep-Q Networks (DQN) is the tendency to incorrectly estimate Q-values. DDQN mitigates this incorrect estimation by using two separate neural networks: an online network of $Q_\theta$ for the action selection and a target network of $Q_{\theta'}$ for $Q$-value evaluation. The update rule for the Q-value is given by
\begin{equation}
Q_\theta(S_t, A_t) \gets Q_\theta(S_t, A_t) + \alpha \left[ Q_t^{\rm targ} - Q_\theta(S_t, A_t) \right],
\label{eq:ddqn_update}
\end{equation}
where $Q_t^{\rm targ}$ is defined as 
\begin{equation}
Q_t^{\rm targ} = R_t + \gamma Q_{\theta'} \left( S_{t+1}, \argmax_{a'} Q_\theta(S_{t+1}, a') \right).
\label{eq:ddqn_target}
\end{equation}
Here, $\alpha$ is the learning rate and $\gamma$ is the discount factor. 
The target network parameters $\theta'$ are synchronized with the online network parameters $\theta$ at regular intervals (every fixed number of episodes). 
We selectively use two neural networks for $Q_\theta$ to select the best action and $Q_{\theta'}$ to evaluate its value, and minimize the difference between the target value $Q_t^{\rm targ}$ and the current Q-value $Q_\theta(S_t, A_t)$, thereby updating the weights and biases of the neural network $Q_\theta$. 
This allows DDQN to mitigate noise-induced estimation errors, resulting in more stable and accurate Q-value predictions.

To further stabilize the training process, we utilize \textit{experience replay} (see Fig.~\ref{exp_replay2}). This mechanism breaks the temporal correlation between consecutive samples by storing transition tuples $(S_t, A_t, R_t, S_{t+1})$ in a replay buffer. During the learning phase, the agent randomly samples mini-batches of $n$-experiences from the buffer to update the weights and biases of $Q_\theta$.
As illustrated in Fig.~\ref{exp_replay2}, this stochastic sampling ensures that the neural network learns from a diverse set of past experiences, preventing it from overfitting to recent trajectories. In our implementation, experience replay is performed at specific step intervals and at the end of each episode, ensuring policy refinement throughout the design process.

\section{Method}
\subsection{Quantum Circuit Design Workflow}

\begin{figure*}[t]
	\centering
    \includegraphics[keepaspectratio, width=120mm]{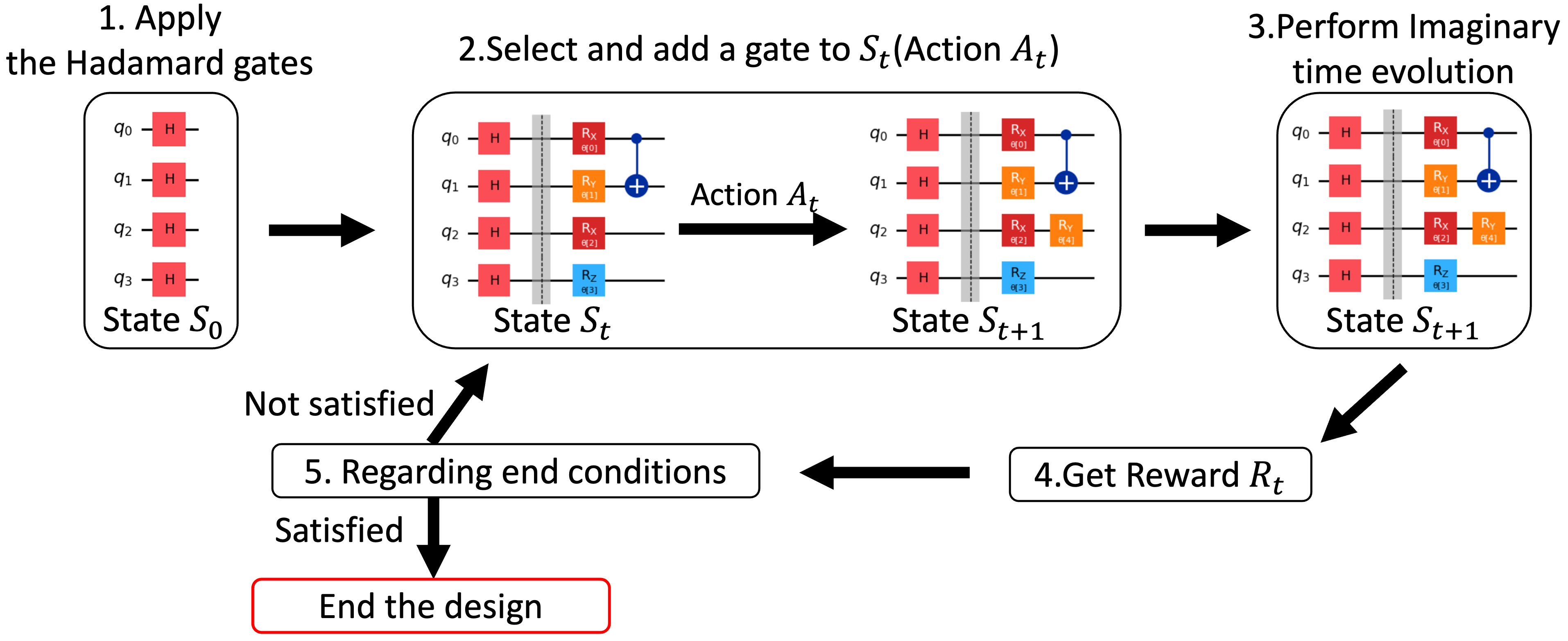}
	\caption{Schematics of Quantum Circuit Design Workflow.}
	\label{design_flow}
\end{figure*}

In this study, we automate the construction of quantum circuits for the VITE. The design process follows a sequential optimization loop, as illustrated in Fig.~\ref{design_flow}. The specific steps are as follows:
\begin{enumerate}
  \item \textbf{Initialization}: An initial circuit $S_0$ is prepared by applying Hadamard gates $(H)$ to all qubits to create a uniform superposition state.
  \item \textbf{Action Selection}: The agent selects an action $A_t$ (adding a gate) to update the current circuit design $S_t$ to $S_{t+1}$.
  \item \textbf{Simulation}: The VITE is performed using the updated circuit $S_{t+1}$. 
  \item \textbf{Reward Calculation}: A scalar reward $R_t$ is computed based on the energy expectation value and the resulting circuit complexity.
  \item \textbf{Termination Check}: The process terminates if predefined conditions are satisfied; otherwise, the agent returns to the step (2).
\end{enumerate}

The design process terminates when one of the following criteria is satisfied: (i) the energy expectation value falls below a predefined threshold $E_{\rm threshold}$ (defined as a \textit{success}), or (ii) the circuit depth or gate count reaches the predefined upper bound (defined as a \textit{failure}). The strategy determining the threshold $E_{\rm threshold}$ will be discussed in Sec.~\ref{E_threshold} and Sec.~\ref{AET}. 

To evaluate the efficiency of our framework, we employ a 4-qubit system for our simulation, which is primarily driven by two factors. 
First, scaling the number of qubits is impractical at this \textit{proof-of-concept} (PoC) stage because the coupling of RL with VITE simulations is computationally intensive, especially given the need for extensive tuning of both hyperparameters and parameters. 
Second, a 4-qubit system for the hydrogen molecule ($H_2$) is the minimal configuration for meaningful quantum chemical simulations. We then employ a 4-qubit hardware-efficient SU$(2)$ ansatz as a baseline~\cite{kandala2017hardware}. 

\begin{figure}[t]
	\centering
    \includegraphics[keepaspectratio, width=87mm]{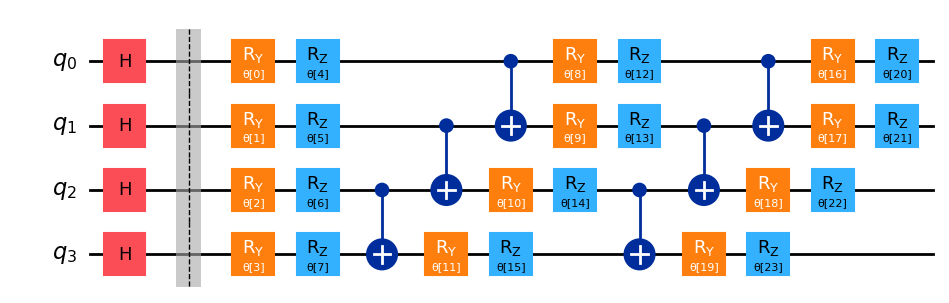}
	\caption{
    Quantum circuit for 4-qubit hardware-efficient SU$(2)$ ansatz (reps=2), referring to the structure following the $H$ gates. The circuit depth is 11 and the gate count is 30; note that the initial $H$ gates are excluded from the count.
    }
	\label{SU2reps2}
\end{figure}

The upper bounds on circuit depth and gate count are determined to maintain an efficient search space while ensuring sufficient expressivity. 
For the evaluation of the proposed framework, the Max-Cut problem serves as a benchmark for combinatorial optimization, and the $H_2$ Hamiltonian is adopted for quantum chemistry applications. 
For the $H_2$ Hamiltonian, the preliminary simulations indicated that a repetition number of $p=1$ in the hardware-efficient SU$(2)$ ansatz is insufficient for ground state convergence, whereas $p=2$ (corresponding to a depth of 11 and 30 gates) is required (Fig.~\ref{SU2reps2}). Accordingly, the agent's upper bounds are set at a depth of 10 and a gate count of 30, excluding the initial Hadamard gates. 
While the gate count bound matches the baseline, the depth bound is intentionally set to be stricter than the 11-layer hardware-efficient SU$(2)$ ansatz. This configuration aims to discover circuit structures that outperform the baseline by achieving higher efficiency with reduced depth for the $H_2$ Hamiltonian. 
Note that since the initialization to add Hadamard gates in the step (1) is a standard procedure in our framework, these gates are excluded from the gate count and circuit depth metrics in this study.

Furthermore, to enhance search efficiency and prune the search space, gates are added column-wise across qubits. 
For example, in a state $S_t$ shown in Fig. \ref{design_flow}, gates are added in the order of $R_x$, $R_y$, $R_x$, $R_z$, and CNOT-gates. 
We further impose two structural constraints:
\begin{itemize}
    \item \textbf{Adjacency Constraint}: Identical single-qubit gates (excluding identity gates) are prohibited from being placed consecutively on the same qubit, even if separated only by identity gates. 
    \item \textbf{Connectivity Constraint}: CNOT gates are restricted to nearest-neighbor qubits, with the control bit at $q_i$ and the target bit at $q_{i+1}$, to adhere hardware efficiency and avoid costly SWAP overhead.
\end{itemize}
Regarding the second constraint, the qubit $q_3$ in Fig. \ref{design_flow} cannot serve as a control qubit for this configuration, since the target bit $q_4$ is not present in the system.

\subsection{Hamiltonian and Problem Instances}\label{HPI}

To evaluate the performance of the proposed automated design framework, we employ two distinct benchmarks: the Max-Cut problem and the ground-state simulation of the hydrogen molecule ($H_2$).

\begin{figure}[t]
    \centering
    \includegraphics[keepaspectratio, width=60mm]{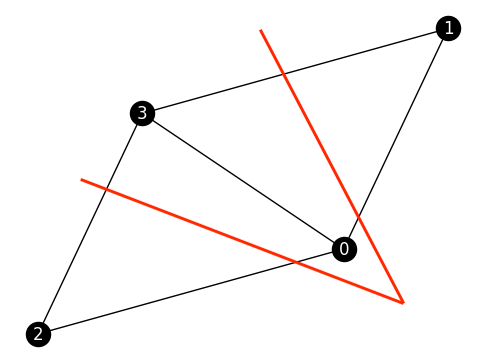}
    \caption{Schematics of the Max-Cut problem. A graph with 4 vertices and its maximum cut. All edge weights are set to 1.}
    \label{Maxcutgraph}
\end{figure}

The Max-Cut problem \cite{goemans1995improved} seeks to partition the vertex set $V$ of an undirected graph $G=(V, E)$ into two disjoint subsets $S$ and $\bar{S}$ such that the sum of the weights of the edges between them is maximized.
In this study, we utilize a 4-vertex graph with uniform edge weights, as illustrated in Fig.~\ref{Maxcutgraph}. The Hamiltonian for this problem is mapped onto a 4-qubit system and consists entirely of diagonal Pauli-$Z$ operators, providing a baseline for evaluating the agent's ability to minimize cost functions in a discrete combinatorial landscape.

For quantum chemical calculations, we target the ground-state energy of the hydrogen molecule ($H_2$). While the $H_2$ system is the simplest molecular model, it serves as a critical PoC for our framework. Specifically, its well-understood energy landscape allows for an unambiguous evaluation of the RL agent's ability to achieve the Full Configuration Interaction (Full-CI) result with minimal circuit depth. By utilizing this fundamental model, we can strictly verify whether the DDQN can eliminate redundant gates that are often present in manually designed or heuristic ansatz.

To transform the fermionic Hamiltonian into a qubit-compatible format, we adopt the Bravyi-Kitaev (BK) transformation \cite{seeley2012bravyi}. 
In contrast to the Jordan-Wigner transformation \cite{fradkin1989jordan}, the BK method maps fermion operators to Pauli strings with a logarithmic scaling of locality, $O(\log N)$, for the $N$-qubit system. This logarithmic mapping effectively reduces the circuit overhead required to achieve chemical accuracy. 
Furthermore, while the Max-Cut Hamiltonian is composed exclusively of diagonal Pauli-$Z$ terms, the $H_2$ Hamiltonian incorporates off-diagonal terms involving Pauli-$X$ and $Y$ operators. These off-diagonal elements induce a complex matrix structure of the Hamiltonian compared with the Max-Cut problem, and the RL agent consequently must explore a broader manifold of the Hilbert space, requiring a more sophisticated and expressive ansatz circuit.

\subsection{Neural Network Architecture}

\begin{figure}[t]
    \centering
    \includegraphics[keepaspectratio, width=85mm]{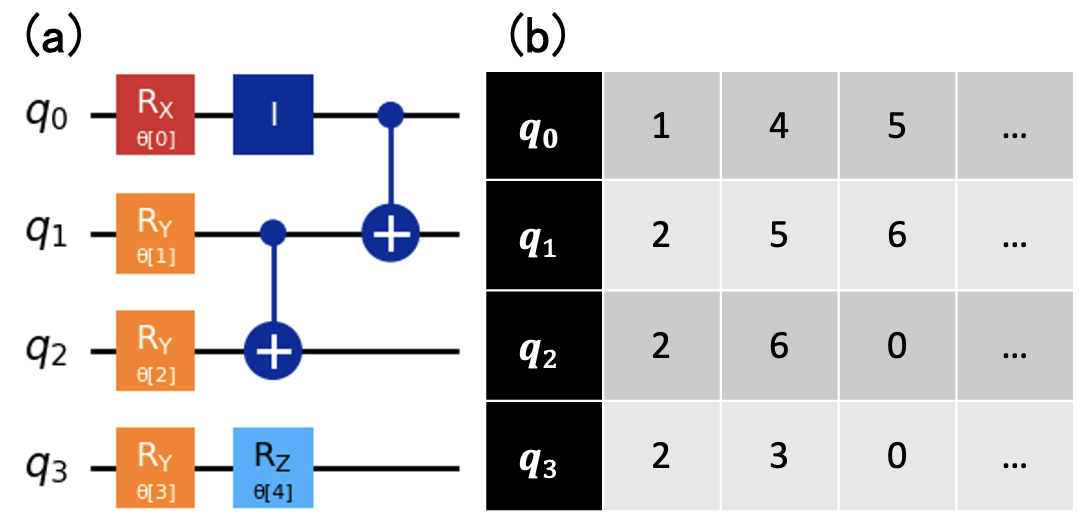}
	\caption{(a) Example of 4-qubit quantum circuit, (b) List representation of the quantum circuit on (a). In the table (b), the label $0$ indicates the absence of the gate, $(1,2,3, 4)$ represents $(R_x, R_y, R_z, I)$, and $(5, 6)$ represents (control, target) part of the CNOT gate.}
    \label{qc_list}
\end{figure}

The DDQN agent utilizes a deep neural network to approximate the action-value function $Q(S_t, A_t)$. The network architecture consists of an input layer, three fully connected (dense) hidden layers, and an output layer. The input to the network is a vectorized representation of the current quantum circuit $S_t$. Specifically, the $4 \times 10$ circuit grid is mapped into a one-dimensional list, where each gate configuration is encoded as integers as illustrated in Fig.~\ref{qc_list}: 
\begin{itemize}
    \item \textbf{0}: Empty slot
    \item \textbf{1--3}: Single-qubit rotation gates ($R_x, R_y, R_z$)
    \item \textbf{4}: Identity gate ($I$)
    \item \textbf{5, 6}: CNOT gate (control and target qubits, respectively)
\end{itemize}

Each of the three hidden layers contains 32 nodes equipped with the Rectified Linear Unit (ReLU) activation function. The output layer consists of five nodes, each corresponding to the estimated Q-value for the available actions for $(R_x,R_y,R_z,I,{\rm CNOT})$. To ensure robust weight updates, we employ the Huber loss function, which is less sensitive to outliers in Q-value estimation than the mean squared error. The network parameters are optimized using the Adam optimizer.

In our implementation, the two networks $Q_\theta$ and $Q_{\theta'}$ are synchronized by copying the weights of the online network to the target network every 30 episodes. This periodic update stabilizes the learning process by providing a consistent target for the Q-value estimation.

\subsection{Detailed Settings}

\subsubsection{Reward Function}

The reward $R_t$ is designed to prioritize both the minimization of the energy expectation value and the compactness of the circuit structure. 
We first employ the following reward: 
\begin{equation}\label{eq:reward}
  R_t = (E_{t-1} - E_t) + c (g_{\max} - g)  \Theta(E_{\text{threshold}} - E_t),
\end{equation}
where $E_{t-1}$ and $E_t$ are the energy expectation values before and after the action $A_t$, respectively. The term $g$ denotes the current gate count, and $g_{\max}$ is the predefined upper limit ($g_{\max}=30$ in the current case). 

We utilize the gate count rather than circuit depth in the reward function to provide a finer evaluation of circuit size. 
The second term incorporates the Heaviside step function $\Theta(x)$, which activates an additional smaller-circuit reward $c (g_{\rm max} - g)$ only when the circuit successfully provides the expected value of the Hamiltonian $E_t$ that is lower than the given threshold $E_{\rm threshold}$. We used the weight hyperparameter $c$ to be $0.1$. 

\subsubsection{Evolving Threshold}\label{E_threshold}

We implement an evolving thresholding mechanism for $E_{\text{threshold}}$. 
The threshold is initialized at $0.0$ since the ground state energies in both benchmark models are known to be negative. To drive the agent toward the lower energy state, $E_{\text{threshold}}$ is updated every 10 successful episodes to $E_{\text{best}} - \epsilon$, where $E_{\text{best}}$ is the minimum energy achieved from the start of learning up to the current learning point, and we used $\epsilon = 0.01$.

Furthermore, a lower bound is imposed on the threshold to ensure the target remains within a physically plausible range, where this bound is set to $0.9 E_{\min}$. 
For the current numerical experiments, we assume $E_{\min}$ is given to validate the framework. Strategies for scenarios where $E_{\min}$ is unavailable will be addressed in the Discussion section. 
Here, the reference ground-state energy $E_{\min}$ is set to $-3.0$ for the Max-Cut problem and the Full-CI value ($E_{\text{FCI}} \approx -1.137$ Ha) for the hydrogen molecule.

\subsubsection{Numerical Setup and Hyperparameters}

As mentioned earlier, the design environment is configured for a 4-qubit system with a maximum depth of 10 and a maximum gate count of 30. The learning process is evaluated based on the energy expectation value, gate count, circuit depth, and cumulative reward.

For the DDQN hyperparameters, the learning rate $\alpha$ is set to $0.001$, and the discount factor $\gamma$ is set to $0.99$. Regarding experience replay, the buffer capacity is $50,000$ episodes. At the end of each episode, the network is updated through 64 iterations using mini-batches of $128$ samples randomly drawn from the replay buffer.

The exploration-exploitation trade-off is governed by the $\varepsilon$-greedy policy. We initialized the exploration probability at $\varepsilon = 1.0$ with a decay rate of $\Gamma = 0.985$ and a lower bound of $0.1$. This persistent $10\%$ exploration probability ensures that the agent continues to discover efficient circuit designs and avoids premature convergence to local minima.

The proposed framework was implemented using {Keras} for the neural network architecture, {OpenAI Gym} for the reinforcement learning environment, and {Qiskit} for quantum circuit simulation and expectation value calculation.

\section{Results} \label{result}

\subsection{Max-Cut Problem}
\begin{figure}[t]
    \centering
    \includegraphics[keepaspectratio, width=85mm]{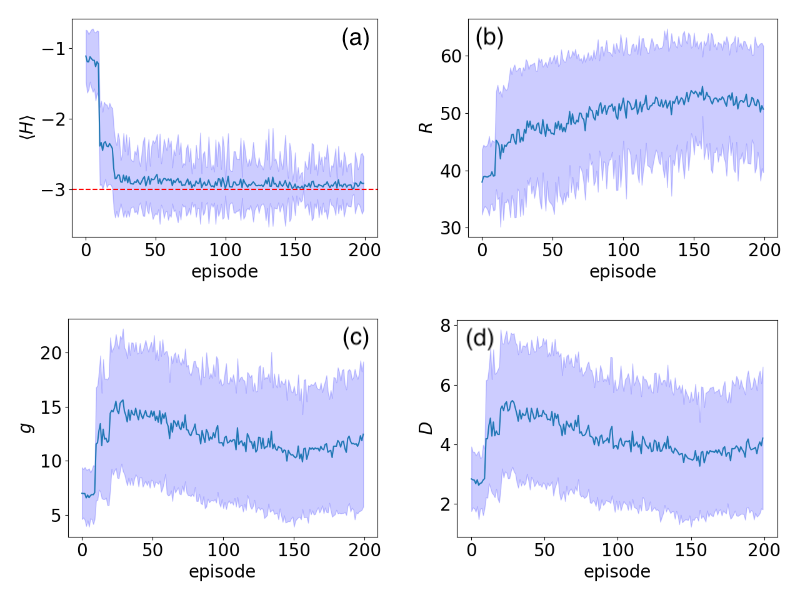}
	\caption{Episode-dependence of (a) the expectation value of the Hamiltonian $\langle H \rangle$, (b) cumulative reward $R$, (c) gate count $g$, and (d) circuit depth $D$ averaged over 100 trials for the Max-Cut problem. The red dashed line in (a) represents the theoretical minimum energy $E_{\rm min}$.}
    \label{max_result}
\end{figure}

The performance of the DDQN agent on the Max-Cut problem over 100 independent trials is summarized in Fig.~\ref{max_result}. As learning progresses, the energy expectation value $\langle H \rangle$ converges toward the theoretical minimum value, and the cumulative reward exhibits a corresponding increase, indicating successful policy optimization. 

Specifically, regarding the circuit complexity, both the average gate count $g$ and circuit depth $D$ exhibit a distinct monotonic decrease from around episode 25 to around episode 150. This interval represents a stable optimization phase where the agent effectively transitions from initial stochastic exploration to a structured design strategy. During this period, the framework consistently identifies and eliminates redundant gates while maintaining the required energy precision.

However, beyond approximately episode 150, we observed a slight reversal in this trend, with both gate count and depth beginning to increase. 
This behavior, accompanied by a decline in cumulative reward, suggests that the agent's policy may become overly sensitive to the increasingly stringent energy requirements of the evolving threshold. 
As $E_{\text{threshold}}$ approaches the global minimum, the agent likely prioritizes marginal energy improvements by introducing compensatory gates, even at the expense of the circuit compactness reward.

This behavior indicates a trade-off between absolute precision and circuit compactness. The instability observed after episode 150 implies that the current reward structure may require further improvement—such as increasing the penalty for circuit complexity—to maintain the monotonic decrease in gate count throughout the entire training duration. 

\begin{figure}[t]
    \centering
    \includegraphics[keepaspectratio, width=40mm]{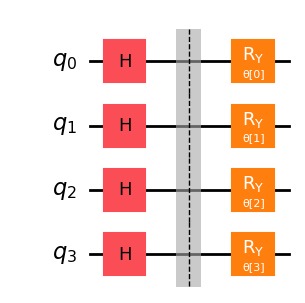}
    \caption{The smallest quantum circuit designed for the Max-Cut problem found in the RL method.}
    \label{qc_min}
\end{figure}

In contrast to the $H_2$ molecular problem, a repetition number of $p=1$ in the hardware-efficient SU$(2)$ ansatz is sufficient for ground state convergence in the Max-Cut problem. Excluding the initial Hadamard gates, this baseline configuration consists of 19 gates and a circuit depth of 7. 
By episode 200, the agent of the RL achieved an average reduction of approximately 37\% in gate count ($g \approx 12$) and 43\% in circuit depth ($D \approx 4$) relative to this hardware-efficient SU$(2)$ baseline with $p=1$. Notably, the proposed framework successfully discovered the optimal circuit ($g=4, D=1$) required to reach $E_{\min} = -3.0$, as illustrated in Fig.~\ref{qc_min}. This optimal configuration represents a significant reduction of 79\% in gate count and 86\% in circuit depth compared to the baseline.

\subsection{Hydrogen Molecular Hamiltonian}
\begin{figure}[t]
    \centering
    \includegraphics[keepaspectratio, width=85mm]{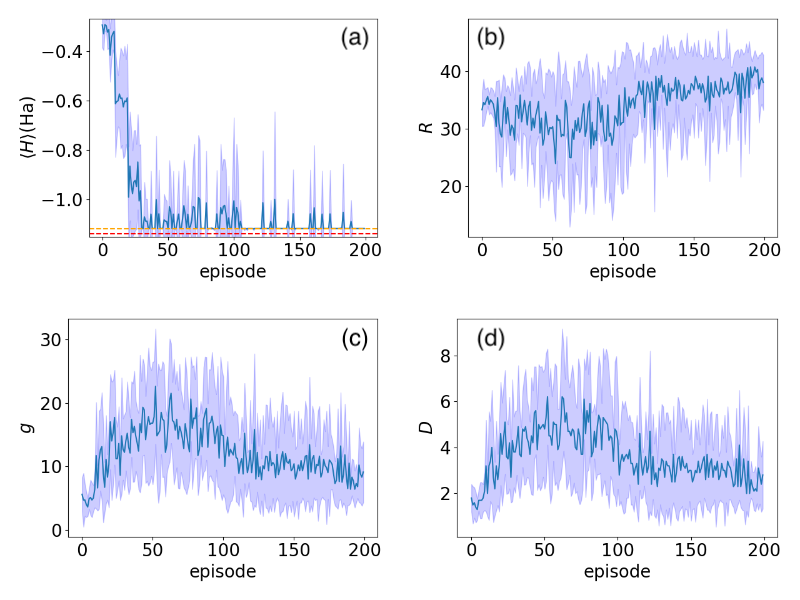}
	\caption{Episode-dependence of (a) the expectation value of the Hamiltonian $\langle H \rangle$, (b) cumulative reward $R$, (c) gate count $g$, and (d) circuit depth $D$ averaged over 100 trials for the hydrogen molecular Hamiltonian. The orange and red lines in (a) represent $E_{\rm HF}$ and $E_{\rm FCI}$, respectively.}
	\label{h2_result}
\end{figure}

The performance of the DDQN agent on the $H_2$ Hamiltonian over 100 trials of 200 episodes is presented in Fig.~\ref{h2_result}. 
Similar to the Max-Cut problem, following an initial intensive exploration phase, we observed a consistent downward trend in the energy expectation value $\langle H \rangle$, average gate count $g$, and circuit depth $D$ as the agent's policy began to stabilize. 
By episode 200, the agent significantly reduced the circuit complexity relative to the hardware-efficient SU(2) baseline with $p=2$ (Fig.~\ref{SU2reps2}), resulting in an average gate count of $g \approx 10$ and a depth of $D \approx 3$, which correspond to reductions of 66\% and 72\%, respectively.

Despite these structural optimizations, the majority of the designed circuits remained near the Hartree-Fock energy level $E_{\text{HF}}$, with only 156 episodes (about 0.78\%) reaching the Full-CI value ($E_{\text{FCI}}$) over 20,000 episodes (100 trials with 200 episodes each).
Unlike the Max-Cut problem, reaching $E_{\min}$ in quantum chemical calculations presents a significantly more complex task; it requires identifying sophisticated gate structures that can both capture essential quantum correlations and maintain high numerical precision. 
Our results suggest that the relatively short training duration of 200 episodes was insufficient for the agent to consistently navigate the complex energy landscape due to the quantum correlation.  

\begin{figure}[t]
    \centering
    \includegraphics[keepaspectratio, width=85mm]{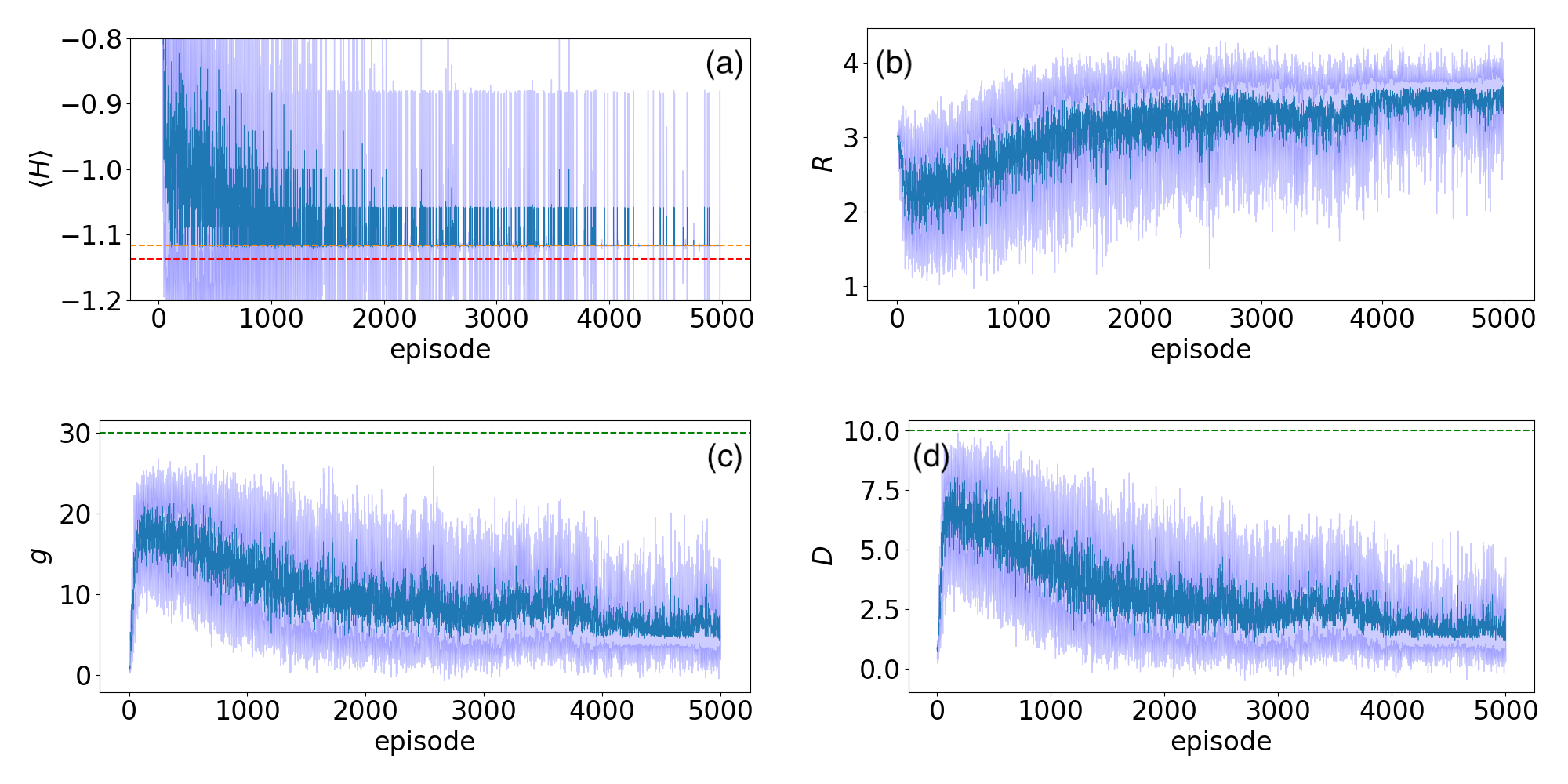}
	\caption{Episode-dependence of (a) the expectation value of the Hamiltonian $\langle H \rangle$, (b) cumulative reward $R$, (c) gate count $g$, and (d) circuit depth $D$ averaged over 10 trials for the hydrogen molecular Hamiltonian upto 5,000 episodes. In (a), the orange and red dashed lines represent $E_{\rm HF}$ and $E_{\rm FCI}$, respectively. In (c) and (d), the green lines denote the prescribed upper bounds. }
	\label{h2_result2}
\end{figure}

To address this, we extended the training duration to 5,000 episodes over 10 independent runs, adjusting the exploration parameters to $\Gamma=0.9993$ and a lower bound of $\varepsilon=0.05$. As shown in Fig.~\ref{h2_result2}, this extended training further streamlined the circuit complexity, achieving an average reduction of 80\% in both gate count ($g \approx 6$) and depth ($D \approx 2$). 
Despite the extended training, only 191 circuits (approximately 0.38\%) reached the Full-CI results within the total 50,000 episodes (10 runs with 5,000 episodes each). 

According to Eq.~\eqref{eq:reward}, if the Hamiltonian expectation value improves from the Hartree-Fock value ($E_{\rm HF} \approx -1.116$) to the Full-CI value ($E_{\rm FCI} \approx -1.137$), the resulting reward increment is approximately $0.021$. 
In contrast, the reduction of a single quantum gate ($g$) yields a reward increase of $c=0.1$ in the current case, which is nearly five times larger than the reward obtained by the recovery of quantum electronic correlation. 
Furthermore, the current threshold for successful circuit design has a lower bound of $0.9E_{\rm FCI} \approx -1.0467$, which is higher than $E_{\rm HF}$. 
This loose constraint allowed the agent to accept mean-field approximations as successes. 
Consequently, the agent first prioritizes falling below the loose threshold higher than $E_{\rm HF}$ before focusing on gate reduction, and then prioritized ``minimizing the gate count" over ``minimizing the energy expectation value" to maximize cumulative rewards. 

\section{Discussion}

The results in the previous section indicate that simply increasing the number of training episodes is insufficient for the agent to converge to the ground state of the hydrogen molecular Hamiltonian. 
This indicates that a more sophisticated reward and threshold structure is vital for attaining the high precision of chemical accuracy.
In the following, we refine the reward structure and the energy threshold mechanism to ensure better precision and optimized circuit complexity.

\subsection{Refined Reward Design}
There are two inherent drawbacks in the reward design used in the previous section. First, it lacks precise normalization across different scales. Second, although $E_{\min}$ is typically unknown in practical applications, it was implicitly incorporated into the reward function as a lower bound of the energy threshold. 

We introduce a reward function $R_t$ that eliminates the need for prior knowledge of $E_{\rm min}$ while ensuring efficient normalization, which is also constructed to balance the competing objectives of energetic optimization and circuit minimality. 
At each step $t$, this reward can be given by 
\begin{equation}\label{reward}
  R_t = \frac{E_{t-1} - E_t}{E_{t-1} - E_{\rm{bound}}} + \frac{g_{\rm{max}} - g}{g_{\rm{max}}} \Theta (E_{\rm{threshold}} - E_t), 
\end{equation}
where $E_{\rm{bound}}$ serves as a theoretical lower bound, derived as the negative sum of the absolute values of the Pauli coefficients $\{\lambda_\alpha\}$ constituting the Hamiltonian $H = \sum \lambda_\alpha P_\alpha$, given by $E_{\rm bound} = - \sum_\alpha |\lambda_\alpha|$~\cite{mateusz}. 
Since the expectation value of any Pauli string $P_\alpha$ satisfies $|\langle P_\alpha \rangle| \leq 1$, this value represents the absolute minimum energy attainable in the operator space. This approach provides a robust normalization constant for the reward signal regardless of whether the true ground state energy is known. For the hydrogen molecular Hamiltonian in this study, the lower bound is calculated as $E_{\rm{bound}} \approx -1.985..$.

The first term in Eq.~\eqref{reward} provides a reward for energy reduction relative to the defined lower bound $E_{\rm{bound}}$. 
The second term, representing the reward for ``circuit efficiency," is activated only if the energy is below the specified threshold $E_{\rm threshold}$.
An episode terminates either when the expectation value $E_t$ falls below the threshold $E_{\rm threshold}$, or when the circuit exceeds the prescribed limits for gate count or depth. At the point of successful termination, the agent receives the bonus $(g_{\rm max} - g)/g_{\rm max}$, which rewards the discovery of compact circuit architectures.

\subsection{Adaptive Energy Thresholding}\label{AET}

To drive the agent toward lower energy states, the threshold $E_{\rm threshold}$ is dynamically updated based on the best performance achieved in earlier episodes. 
Based on the adaptive threshold strategy for the VQE~\cite{mateusz}, we define the threshold to reflect the minimum energy discovered across all preceding episodes as well as to ensure a progressive minimization of the energy expectation value. 

The initial value of the threshold $E_{\rm threshold, 0}$ for the hydrogen molecular Hamiltonian is defined by incorporating an energetic margin $\xi_0$ relative to the Hartree-Fock energy, expressed as $E_{\rm HF} + \xi_0$.
To ensure a continuous optimization drive, the threshold is updated through two concurrent adaptive mechanisms. 

First, the threshold $E_{\rm threshold}$ is recalibrated every 200 episodes to prevent the agent from becoming trapped in local minima and to encourage broader exploration. The update rule is defined as follows:
\begin{equation}\label{threshold_periodic}
  E_{\rm threshold} \leftarrow 
  \begin{cases}
    E_{\rm threshold} + \xi & \text{if } E_{\rm threshold} < E_{\rm best} \\
    E_{\rm best} + \xi      & \text{if } E_{\rm threshold} \ge E_{\rm best}, 
  \end{cases}
\end{equation}
where $E_{\text{best}}$ represents the minimum energy recorded across all training episodes up to the current step, and $\xi > 0$ is a small margin introduced to intentionally relax the constraint. By periodically resetting the threshold to a value slightly higher than the historical minimum $E_{\rm best}$, we provide the agent with the ``room" to explore diverse gate sequences that might temporarily yield higher energy but eventually lead to more efficient circuit architectures. 

In parallel, a performance-driven update is triggered once the agent demonstrates stability, defined as achieving 20 consecutive successful designs at the current threshold. This mechanism ensures that as the agent masters a certain accuracy level, the threshold is further tightened to drive convergence toward the true ground state:
\begin{equation}\label{threshold_incremental}
  E_{\rm threshold} \leftarrow 
  \begin{cases}
    E_{\rm best}              & \text{if } E_{\rm threshold} < E_{\rm best} \\
    E_{\rm threshold} - \epsilon     & \text{if } E_{\rm threshold} \ge E_{\rm best}, 
  \end{cases}
\end{equation}
Consistent with our reward design, the global lower bound for these updates is maintained at $E_{\rm bound}$, the theoretical operator-space limit.
In this study, we employ $\xi_0 = \xi = 0.005$, and $\epsilon = 0.0005$.

\subsection{Performance under Refined Reward and Adoptive Threshold}

\begin{figure}[t]
	\centering
    \includegraphics[keepaspectratio, width=85mm]{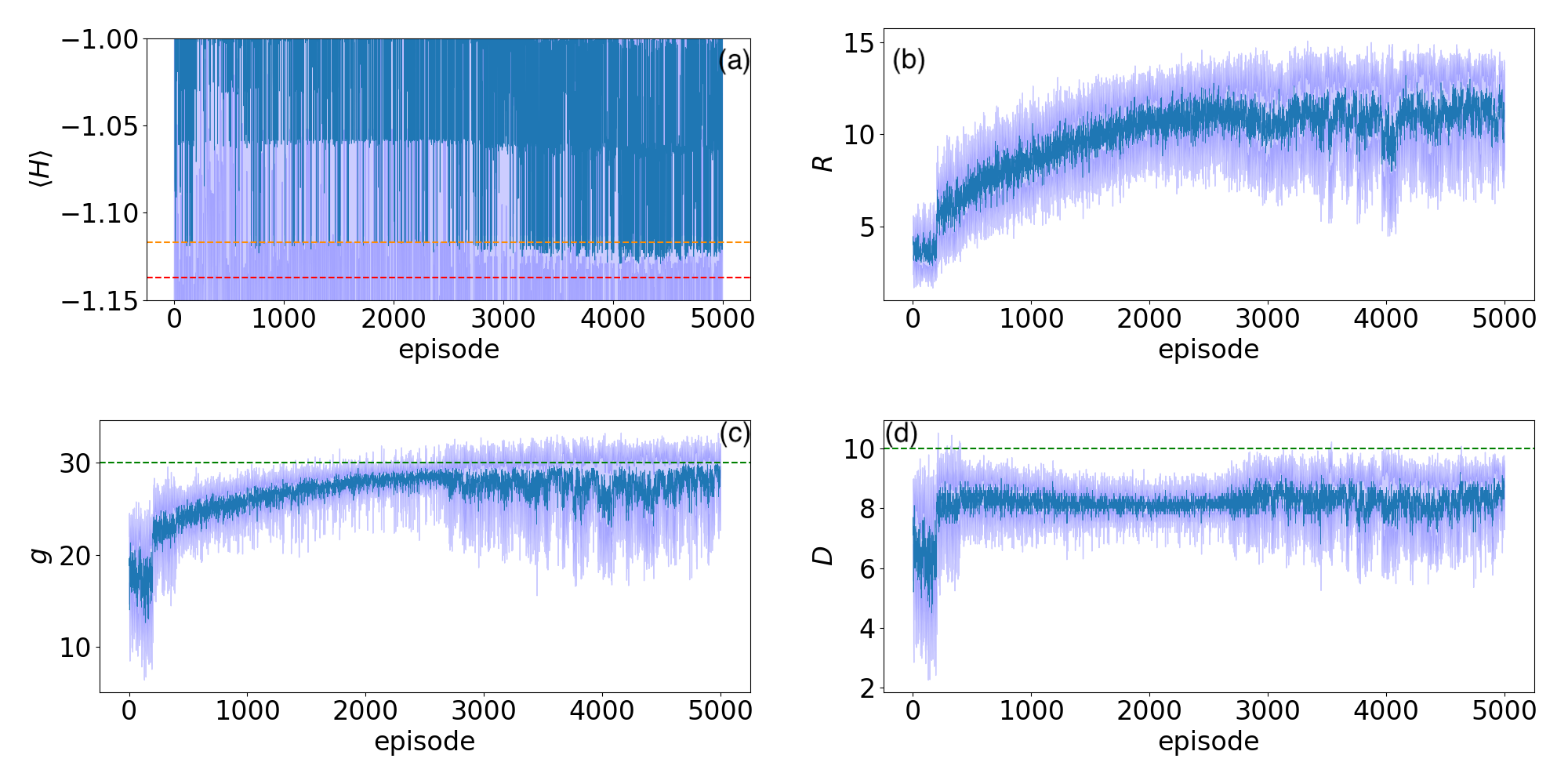}
	\caption{Episode-dependence of (a) the expectation value of the Hamiltonian $\langle H \rangle$(Ha), (b) cumulative reward $R$, (c) gate count $g$, and (d) circuit depth $D$ averaged over 10 trials for the hydrogen molecular Hamiltonian up to 5,000 episodes. We here employ the RL method with the normalized regard and the adaptive energy threshold. In (a), the orange and red dashed lines represent $E_{\rm HF}$ and $E_{\rm FCI}$, respectively. In (c) and (d), the green lines denote the prescribed upper bounds. }
	\label{h2_result3}
\end{figure}

To evaluate the efficacy of the refined reward functional and the adaptive thresholding strategy, we conducted 10 independent trials of 5,000 episodes each for the $H_2$ molecular Hamiltonian. Fig.~\ref{h2_result3} illustrates the evolutionary trajectory of the expectation value of the Hamiltonian, cumulative reward, number of the gates, and the circuit depth. 
The stricter success criteria forced the agent to utilize more optimization steps, causing the gate count and circuit depth to remain near their upper bounds.
Meanwhile, the average expectation value of the Hamiltonian exhibited a downward trend, consistently surpassing the Hartree-Fock value ($E_{\text{HF}}$) around the 3,200-th episode. Notably, the policy successfully resolved the Full-CI energy ($E_{\text{FCI}}$) in 4,458 instances across 50,000 total episodes (10 trials with 5,000 episodes each). This represents a success rate of approximately 8.91\%, a significant order-of-magnitude improvement over the 0.38\% observed in the previous setup. 

For the circuits achieving chemical accuracy, the average gate count and depth were 20.78 and 6.94, respectively. Compared to the conventional hardware-efficient SU$(2)$ ansatz with $p=2$ (Fig.~\ref{SU2reps2}), our method achieved reduction rates of 31\% in gate count and 37\% in circuit depth. The increasing frequency of the circuit that reaches $E_{\text{FCI}}$ solutions beyond 3,000 episodes suggests that extended training further stabilizes the policy toward high-precision regimes in this setup.

While the proposed framework successfully identifies compact circuit architectures, the computational overhead---approximately three weeks to one month for 10 independent runs, each spanning 5,000 episodes, in our numerical environment---suggests that 
performing a full reinforcement learning cycle for each new problem entails significant computational overhead, limiting its feasibility for practical deployments.
Rather than treating this as a real-time problem solver, its strategic value is two-fold. 
First, it serves as a high-quality generator of training datasets for supervised learning models, providing an informative corpus of ``optimal" circuit configurations. 
Second, it facilitates the extraction of universal structural motifs from the obtained quantum circuits, enabling the identification of fundamental gate sequences essential for capturing quantum correlation. 
We expect these insights to be a crucial step toward establishing automated design guidelines for quantum circuit ansatz that transcend individual problem instances. 

\begin{figure}[t]
    \centering
    \includegraphics[keepaspectratio, width=85mm]{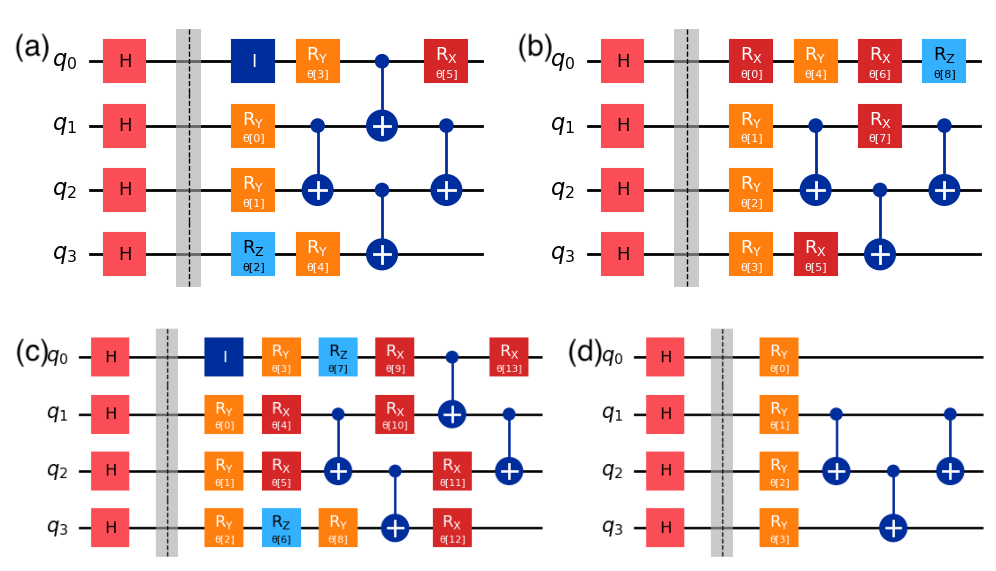}
	\caption{(a)-(c) Examples of circuits reaching $E_{\rm FCI}$ obtained in the present RL method. By comparing the architectures of circuits (a)–(c) and isolating their overlapping components, the ``essential skeleton" was identified in circuit (d); this refined structure successfully reached $E_{\text{FCI}}$.}
    \label{h2_min}
\end{figure}

For the second point, while the obtained circuits reaching $E_{\text{FCI}}$ exhibit various gate patterns (Fig.~\ref{h2_min}(a)-(c)), further structural analysis allows for the isolation of their most critical components. 
By identifying and extracting these core functional motifs while eliminating redundant operations, we arrive at the streamlined circuit shown in Fig.~\ref{h2_min}(d) that can also provides the $E_{\text{FCI}}$ result for the VITE. 
This configuration represents a significant reduction of 77\% in gate count ($g=7$) and 64\% in circuit depth ($D=4$) compared to the baseline. 
This reduction from (a)–(c) to (d) demonstrates that while there is still scope for further refinement in our RL framework—such as incorporating the adaptive threshold strategy to not only the expectation value of the Hamiltonian but also gate count—the generated circuits effectively reveal the ``essential skeleton" required for high accuracy.

\section{Conclusion}

In this study, we developed an automated quantum circuit design framework for the variational imaginary time evolution (VITE) method using double Deep-Q Networks (DDQN). Through evaluations of the Max-Cut problem and the hydrogen molecular Hamiltonian, we demonstrated that the reinforcement learning (RL) agent can generate the quantum circuits that can successfully optimize both energy and circuit complexity. 

For applications requiring computational accuracy, such as quantum chemistry calculations, the results tend to be approximate solutions if we choose a loose energy threshold. 
By refining the reward function and implementing an adaptive thresholding mechanism, we significantly improved the agent's ability to reach chemical accuracy. 
Additionally, we demonstrated that extracting common functional motifs from these RL-optimized circuits facilitates the identification of a ``essential skeleton." 
This finding suggests that the proposed method serves as both a high-quality circuit generator of training data and a powerful heuristic for establishing universal design guidelines for compact, problem-specific quantum ansatz.

Future work will focus on scaling this framework to larger systems and investigating its effectiveness across diverse quantum chemistry, combinatorial optimization and quantum machine learning problems. 
A key refinement will involve extending the adaptive threshold strategy to optimize not only the Hamiltonian expectation value but also the gate count.
Furthermore, the establishment of design guidelines tailored to specific problem properties will be facilitated by building a comprehensive corpus of optimal quantum circuits, derived from the automated design of diverse systems. 

\section*{Acknowldgement}
We thank Kota Mizuno for fruitful discussions. This study was supported by JST PRESTO JPMJPR211A and the New Energy and Industrial Technology Development Organization (NEDO, JPNP25014).

\bibliography{ebib}

@article{preskill2018quantum,
  title={Quantum computing in the NISQ era and beyond},
  author={Preskill, John},
  journal={Quantum},
  volume={2},
  pages={79},
  year={2018},
  publisher={Verein zur F{\"o}rderung des Open Access Publizierens in den Quantenwissenschaften}
}

@article{mateusz,
  title = {Reinforcement learning for optimization of variational quantum circuit architectures},
      author={Mateusz Ostaszewski and Lea M. Trenkwalder and Wojciech Masarczyk and Eleanor Scerri and Vedran Dunjko},
  journal = {Advances in Neural Information Processing Systems},
  volume = {34},
  pages = {18182--18194},
  year = {2021}
}

@article{fosel,
  title={Quantum circuit optimization with deep reinforcement learning},
  author={F{\"o}sel, Thomas and Niu, Murphy Yuezhen and Marquardt, Florian and Li, Li},
  journal={arXiv preprint arXiv:2103.07585},
  year={2021}
}

@article{kolle2024reinforcement,
  title={A reinforcement learning environment for directed quantum circuit synthesis},
  author={K{\"o}lle, Michael and Schubert, Tom and Altmann, Philipp and Zorn, Maximilian and Stein, Jonas and Linnhoff-Popien, Claudia},
  journal={arXiv preprint arXiv:2401.07054},
  year={2024}
}

@article{kimura2022quantum,
  title={Quantum circuit architectures via quantum observable markov decision process planning},
  author={Kimura, Tomoaki and Shiba, Kodai and Chen, Chih-Chieh and Sogabe, Masaru and Sakamoto, Katsuyoshi and Sogabe, Tomah},
  journal={Journal of Physics Communications},
  volume={6},
  number={7},
  pages={075006},
  year={2022},
  publisher={IOP Publishing}
}

@article{cerezo2021variational,
  title={Variational quantum algorithms},
  author={Cerezo, Marco and Arrasmith, Andrew and Babbush, Ryan and Benjamin, Simon C and Endo, Suguru and Fujii, Keisuke and McClean, Jarrod R and Mitarai, Kosuke and Yuan, Xiao and Cincio, Lukasz and others},
  journal={Nature Reviews Physics},
  volume={3},
  number={9},
  pages={625--644},
  year={2021},
  publisher={Nature Publishing Group UK London}
}

@article{kandala2017hardware,
  title={Hardware-efficient variational quantum eigensolver for small molecules and quantum magnets},
  author={Kandala, Abhinav and Mezzacapo, Antonio and Temme, Kristan and Takita, Maika and Brink, Markus and Chow, Jerry M and Gambetta, Jay M},
  journal={nature},
  volume={549},
  number={7671},
  pages={242--246},
  year={2017},
  publisher={Nature Publishing Group}
}

@article{farhi2014quantum,
  title={A quantum approximate optimization algorithm},
  author={Farhi, Edward and Goldstone, Jeffrey and Gutmann, Sam},
  journal={arXiv preprint arXiv:1411.4028},
  year={2014}
}

@article{schulman2017proximal,
  title={Proximal policy optimization algorithms},
  author={Schulman, John and Wolski, Filip and Dhariwal, Prafulla and Radford, Alec and Klimov, Oleg},
  journal={arXiv preprint arXiv:1707.06347},
  year={2017}
}

@article{motta2020determining,
  title={Determining eigenstates and thermal states on a quantum computer using quantum imaginary time evolution},
  author={Motta, Mario and Sun, Chong and Tan, Adrian TK and O’Rourke, Matthew J and Ye, Erika and Minnich, Austin J and Brandao, Fernando GSL and Chan, Garnet Kin-Lic},
  journal={Nature Physics},
  volume={16},
  number={2},
  pages={205--210},
  year={2020},
  publisher={Nature Publishing Group UK London}
}

@article{yuan2019theory,
  title={Theory of variational quantum simulation},
  author={Yuan, Xiao and Endo, Suguru and Zhao, Qi and Li, Ying and Benjamin, Simon C},
  journal={Quantum},
  volume={3},
  pages={191},
  year={2019},
  publisher={Verein zur F{\"o}rderung des Open Access Publizierens in den Quantenwissenschaften}
}

@article{mclachlan1964variational,
  title={A variational solution of the time-dependent Schrodinger equation},
  author={McLachlan, Andrew D},
  journal={Molecular Physics},
  volume={8},
  number={1},
  pages={39--44},
  year={1964},
  publisher={Taylor \& Francis}
}

@article{broeckhove1988equivalence,
  title={On the equivalence of time-dependent variational principles},
  author={Broeckhove, J and Lathouwers, L and Kesteloot, E and Van Leuven, P},
  journal={Chemical physics letters},
  volume={149},
  number={5-6},
  pages={547--550},
  year={1988},
  publisher={Elsevier}
}

@article{wick1954properties,
  title={Properties of Bethe-Salpeter wave functions},
  author={Wick, Gian-Carlo},
  journal={Physical Review},
  volume={96},
  number={4},
  pages={1124},
  year={1954},
  publisher={APS}
}

@article{kaelbling1996reinforcement,
  title={Reinforcement learning: A survey},
  author={Kaelbling, Leslie Pack and Littman, Michael L and Moore, Andrew W},
  journal={Journal of artificial intelligence research},
  volume={4},
  pages={237--285},
  year={1996}
}

@inproceedings{tokic2010adaptive,
  title={Adaptive $\varepsilon$-greedy exploration in reinforcement learning based on value differences},
  author={Tokic, Michel},
  booktitle={Annual conference on artificial intelligence},
  pages={203--210},
  year={2010},
  organization={Springer}
}

@inproceedings{van2016deep,
  title={Deep reinforcement learning with double q-learning},
  author={Van Hasselt, Hado and Guez, Arthur and Silver, David},
  booktitle={Proceedings of the AAAI conference on artificial intelligence},
  volume={30},
  number={1},
  year={2016}
}

@article{goemans1995improved,
  title={Improved approximation algorithms for maximum cut and satisfiability problems using semidefinite programming},
  author={Goemans, Michel X and Williamson, David P},
  journal={Journal of the ACM (JACM)},
  volume={42},
  number={6},
  pages={1115--1145},
  year={1995},
  publisher={ACM New York, NY, USA}
}

@article{seeley2012bravyi,
  title={The Bravyi-Kitaev transformation for quantum computation of electronic structure},
  author={Seeley, Jacob T and Richard, Martin J and Love, Peter J},
  journal={The Journal of chemical physics},
  volume={137},
  number={22},
  year={2012},
  publisher={AIP Publishing}
}

@article{fradkin1989jordan,
  title={Jordan-Wigner transformation for quantum-spin systems in two dimensions and fractional statistics},
  author={Fradkin, Eduardo},
  journal={Physical review letters},
  volume={63},
  number={3},
  pages={322},
  year={1989},
  publisher={APS}
}

\end{document}